\documentclass[10pt]{iopart}
\pdfoutput=1
\usepackage{graphicx}
\usepackage{setstack}
\usepackage{iopams}
\usepackage[T1]{fontenc}
\usepackage[latin9]{inputenc}
\usepackage{bm}
\usepackage{babel}
\usepackage{color}
\usepackage{comment}
\usepackage{hyperref}

\makeatletter
\usepackage{babel}

\makeatother

\usepackage{babel}
\begin{document}
\title{The polarizability of a confined atomic system: An application of
Dalgarno-Lewis method}
\author{T. V. C. Antão$^{1}$ and N. M. R. Peres$^{1,2}$}
\address{$^{1}$Centro de Física and  Departamento de Física,  Universidade
do Minho, P-4710-057 Braga, Portugal}
\address{$^{2}$International Iberian Nanotechnology Laboratory (INL), Av Mestre
José Veiga, 4715-330 Braga, Portugal}
\begin{abstract}
In this paper we give an application of Dalgarno-Lewis method,  the
latter not usually taught in quantum mechanics courses. This is very
unfortunate since this method allows to bypass the sum over states
appearing in the usual perturbation theory. In this context, and as
an example, we study the effect of an external field, both static
and frequency dependent, on a model-atom at fixed distance from a
substrate. This can happen, for instance, when some organic molecule
binds from one side to the substrate and from the other side to an
atom or any other polarizable system. We model the polarizable atom
by a short range potential, a Dirac$-\delta$ and find that the existence
of a bound state depends on the ratio of the effective ``nuclear
charge'' to the distance of the atom to the substrate. Using an asymptotic
analysis, previously developed in the context of a single $\delta-$function
potential in an infinite medium, we determine the ionization rate
and the Stark shift of our system. Using Dalgarno-Lewis theory we
find an exact expression for the static and dynamic polarizabilities
of our system valid to all distances. We show that the polarizability
is extremely sensitive to the distance to the substrate creating the
possibility of using this quantity as a nanometric ruler. Furthermore,
the line shape of the dynamic polarizability is also extremely sensitive
to the distance to the substrate, thus providing another route to
measure nanometric distances.  The ditactic value
of the $\delta-$function potential is well accepted in teaching activities
due to its simplicity, while keeping the essential ingredients of
a given problem.
\end{abstract}
\maketitle

\section{Introduction}

\begin{figure}[h]
\begin{centering}
\includegraphics[width=8cm]{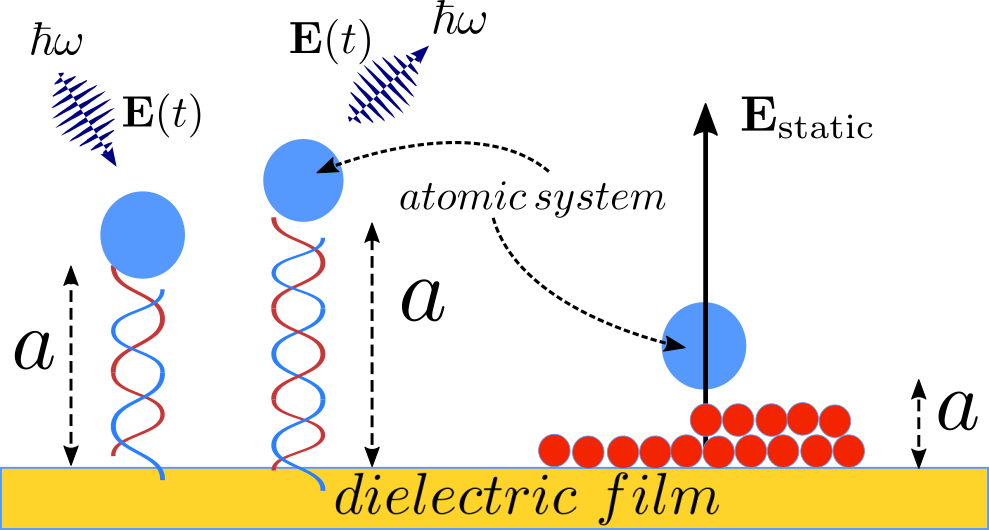}
\par\end{centering}
\caption{Schematic illustration of organic chains with length $a$ bound to
an atomic system (blue dots) at one end and to a dieletric substrate
at the other. The length of the organic chain allows for the control
of the distance of the atomic system to the dieletric substrate. When
an external frequency-dependent electric field is applied to the system,
the dynamic polarizability can be modified by changing the organic
molecule length, which results in the creation of electromagnetic
hot-spots and therefore in the enhancement of Raman-scattering signals,
allowing for detection. In addition, for smaller distances (see right
hand side of the figure with one and two layers of atoms (red dots)
separating the atomic system from the substrate), as will be argued
in the following sections, the static polarizability depends very
sensitively on the the distance $a$, and as such, by measuring this
polarizability, one can obtain information about the number of atomic
layers separating the atomic system from the substrate . This system
can thus be used as a nanometric ruler.\label{fig:Schematic-illustration-of}}
\end{figure}

It is well known that the electric and optical properties of atomic
systems are influenced by the presence of interfaces as discussed
 \cite{chaen2005}. In particular, the response to an electric
field, characterized by shifts of the atom (or molecule) energy levels,
known as the Stark shift, is no exception as shown in \cite{Flatte2008}.
 The Stark shift is often presented in a second course
in quantum mechanics in the context of perturbation theory, however,
this discussion seldom comes with an emphasis on a very common problem
in perturbative calculations, which is the need to sum over an infinite
number (or even a continuum) of states to evaluate the second order
correction. This problem is exacerbated when a non-trivial geometry
is considered, such as the mentioned proximity of the atomic system
to an interface.   In addition to the Stark shift, the energy levels,
which are bound states in the absence of the applied field, become
resonances and an ionization rate can also be defined from a complex
energy eigenvalue. Physically, the electrons escape from the atom
or molecule by tunneling through the finite barrier created by the
applied electric field. It is often of interest to evaluate the electric
response of such atomic or molecular systems as a function of distance
to the interface. For instance, there exist relevant applications
in biophysics related to the detection of organic molecules utilizing
methods of Surface-Enhanced Raman Spectroscopy (SERS), discussed in
\cite{Braun,gkl}, as well as other methods of probing their binding
configuration, for instance those described in \cite{Berlanga_2017}.
The Raman-effect, in particular, can be strongly enhanced by attaching
artificial atoms to one end of organic chains. In such a system, organic
molecules will bind to a dielectric substrate at the one end and to
an atomic system or quantum dot at the other, as illustrated in Fig.
\ref{fig:Schematic-illustration-of}. In the same figure, we also
illustrate the use of the static polarizability to measure the number
of deposited atomic layers on a substrate (right part of the figure).

Varying the distance $a$ (see Fig. \ref{fig:Schematic-illustration-of})
leads to significant changes in the dynamic polarizability of the
atomic system and, consequently, to the optical scattering cross section
of electromagnetic radiation (which is proportional to the imaginary
part of the dynamic polarizability as discussed in \cite{Hulst1958,Hulst1981}).
These changes to the polarizability, in turn, may result in the creation
of electromagnetic hot-spots which result in the amplification of
the Raman-scattering signal of the organic molecules.

We also note that there exits, as will be argued in the next sections,
a strong dependence of the static polarizability on $a$. We are,
thus, driven to suggest that measurements of the static polarizability
of such a system may act as a kind of nanometric ruler which allows
for the determination of the thickness of atomic layers. To quantitatively
describe the dependence of the polarizability on $a$, we consider
a simple 1D model for an hydrogen-like atom. The model consists of
a short range interaction given by a $\delta$-function to represent
the atomic system.   The $\delta$-function is often
used as a toy-model for atomic systems since it manages to keep the
mathematical complexity at a relative minimum while still providing
insight into their qualitative and quantitative behavior. For instance,
in the classic textbook by ohen-Tannoudji {\it et al. } \cite{Cohen-Tannoudji:101367}, an example
of two $\delta-$functions being used to model a molecule is explored
and manages to capture the essential aspects of the bonds formed between
atoms. Nevertheless, this simplified approach has been shown to provide
accurate results in the prediction of the electrical properties of
3D $H_{2}^{+}$ molecules in \cite{Henriques2021EJP}, despite being
of considerably more simplicity than the full Coulomb problem.

In the particular case of the system we are considering,
this approach has the advantage of allowing analytic solutions to
the quantities relevant to our calculations, such as the static and
dynamic polarizability of the atomic system, which makes it a pedagogically
valuable model for introducing perturbative calculations and keeping
the focus away from potential mathematical or numerical details of
the description of the unperturbed system.

In our study, we will use asymptotic methods for obtaining the ionization
rate and Stark shift as well as the polarizability when such a system
is subjected to an external and static electric field. This procedure,
developed in \cite{Fernandez} relies on the asymptotic expansions
of the Airy functions (other methods exist, such that developed in
\cite{Geltman}). These are exact solutions to the Schrödinger equation
in the presence of a static electric field and allows for expressions
to be found for both ionization rate and Stark shift when the distance
from the atomic system to the substrate is small. Emphasis will be
given, however, to the method developed in \cite{Dalgarno,Karplus1962,karplus_1963,mossman_2016}
which is useful in replacing sums over an infinite number of states,
often impossible to calculate directly, by the solution of an inhomogeneous
differential equation. This method allows us to provide an analytic
expression for the static polarizability of the atomic system as a
function of distance to the surface for an arbitrarily large or small
value of this quantity.

In addition, we provide a means of calculating the dynamic polarizability,
which characterizes the response of the atomic system to a frequency-dependent
electric field using a method developed in \cite{Fowler}.  
This method is a very natural extension of Dalgarno-Lewis theory to
time and frequency dependent perturbations. This stress the didactic
value of the method for tackling time-dependent problems.

 We would like to remark, yet again, that the method
of Dalgarno-Lewis is very powerful for bypassing the sum over states
appearing in the usual perturbation theory. This makes it a valuable
tool when going beyond the simplest examples discussed in quantum
mechanics courses where this sum is solvable analytically or often
reduced to a few terms.  This paper presents an application of this
method in the context of a relatively simple quantum-mechanical model
which has a certain degree of physical relevance as pointed out above.
The basic model is a simple problem in quantum mechanics, but the
application of an ac field or a dc field makes this problem non-trivial.
 We believe that the problem we tackle in this paper
has a high pedagogical value for advanced undergraduate students in
quantum mechanics or students starting their graduate studies, as
well as managing to provide insight into the physics that underlie
the model.

This paper is organized as follows: in Sec. \ref{sec:The-ionization-rate}
we introduce our model and give the asymptotic solution of the complex
eigenvalue problem. In Sec. \ref{sec:Exact-calculation-of} we provide
the exact solution to both the static and the dynamic polarizabilities
of our system, using Dalgarno-Lewis' method. In Sec. \ref{sec:Conclusion-and-outlook}
we provide our conclusions. The paper ends with an appendix on the
solution of the inhomogeneous differential equation.

\section{The ionization rate using Fernández's method\label{sec:The-ionization-rate}}

In this section we introduce our model and obtain the condition for
the existence of a bound state. Later, using asymptotic methods we
derive the ionization rate and the Stark shift for our system when
subjected to an external and static electric field.

\subsection{Model and condition for the existence of a bound state}

We are interested in obtaining the shift in energy levels as well
as ionization rate of an atomic system in proximity to a substrate.
The problem, therefore, amounts to solving a Schrödinger equation
with total potential energy given by the sum of the attractive nuclear
potential close to a substrate and the perturbation caused by a static
electric field. We note, however, that the solution of the Schrödinger
equation with Hamiltonian given by the Coulomb potential is cumbersome,
if not impossible for this geometry. As such, we use a very simple
model consisting of a single attractive $\delta$-function potential
energy. It has been noted before in the literature, that the delta
function is a good one-dimensional model for the three dimensional
atom with the full Coulomb interaction. This similarity arises from
the fact that both potentials are singular at the origin as well as
behaving in a similar fashion under derivatives, that is we have that
$\delta'(x)=-\delta(x)/x $ and $\nabla1/r=-1/r^{2}$,
which corresponds to the same functional relation for both potentials
and implies the same form of the virial equation (see \cite{Fernandez,Henriques2021EJP}).
 This potential therefore allows us to keep the mathematics
relatively simple, while capturing the essential physical features
of our system. For this reason, the $\delta-$function model is useful
for obtaining a rough insight into the physics atomic-line systems
and can often be used as a first introduction to atomic and molecular
dynamics in an introductory quantum mechanics course. We shall see
that the solution to the Schrödinger equation for our system in the
absence of the applied electric field is straightforward. Written
in atomic units, such that $\hbar=e=m=1$, the Schrodinger equation
is given by

\begin{equation}
\left(-\frac{1}{2}\frac{d^{2}}{dx^{2}}-Z\delta(x)-\varepsilon_{b}\right)u_{b}(x)=0.\label{eq:schro}
\end{equation}
The solution of the Schrödinger equation will then give the bound
state wave-function $u_{b}$ with bound state energy $\varepsilon_{b}$.
The parameter $Z$ represents an effective nuclear charge. The presence
of the substrate at a distance $x=-a$ imposes a boundary condition
for the wave function such that $u_{b}(-a)=0$. Since we are interested
in bound states, we consider only states with negative energy, written
as $\varepsilon_{b}=-k_{b}^{2}/2$. For $x\neq0$ the solution to
the equation is given as a linear combination of exponentials with
positive and negative arguments. As such, to fulfill the boundary
condition at $x=-a$ and to obtain a normalizable wave function we
must have an evanescent wave for $x>0$ and a hyperbolic sine for
$-a<x<0$. The coefficients can be obtained imposing the continuity
of the wave function at $x=0$. This gives the full solution of the
bound state wave-function

\begin{equation}
u_{b}(x)=\begin{cases}
0  x<-a,\\
2Ae^{-k_{b}a}\sinh(k_{b}(x+a))  -a<x<0\\
2Ae^{-k_{b}a}\sinh(k_{b}a)e^{-k_{b}x}  x>0.
\end{cases},\label{eq:gs}
\end{equation}

\begin{figure}[h]
\begin{centering}
\includegraphics[scale=0.4]{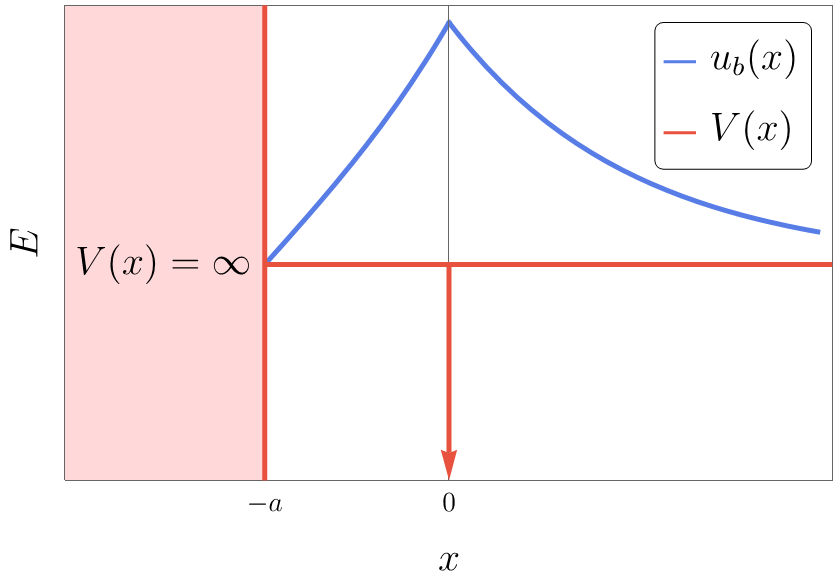}
\par\end{centering}
\caption{Illustration of atomic model potential $V(x)$ (red) and bound state
wave function $u_{b}(x)$ (blue) obtained by the described procedure.
The region coloured in light red corresponds to the substrate which
we model by an impenetrable surface with an infinite potential. The
remaining part of the potential corresponds to the short range Dirac-$\delta$
which goes to $-\infty$ at $x=0$ and is zero otherwise. The wave
function is a hyperbolic sine for $0<x<-a$, and a decaying exponential
for $x>0$. \label{fig:Illustration-of-atomic}}
\end{figure}

Here, the normalization coefficient is given by the radical $A=\sqrt{k_{b}/(1-e^{-2k_{b}a}(1+2k_{b}a)}$.
For further physical insight into the shape and behavior of the wave
function, we plot it, along with the potential in Fig. \ref{fig:Illustration-of-atomic}
.

To obtain the bound state energy we look at the discontinuity of the
derivative of the wave function characteristic of delta-function potentials
$u_{b}'(0^{+})-u_{b}'(0^{-})=-2Zu_{b}(0)$. This allows us to calculate
the ``wave vector'' $k_{b}$ as the solution to a transcendental
equation, which can be easily be obtained numerically, and reads

\begin{equation}
\frac{k_{b}}{Z}=1-e^{-2k_{b}a}.\label{eq:ktrans}
\end{equation}

We note that as the distance from the model atom to the barrier goes
to zero, we obtain $1/Z=2a(1+k_{b}a)$. This equation can only have
a solution if $Z>1/2a$. Therefore we have a limiting condition for
the existence of a bound state in our problem. Note that for the isolated
model atom, where the wave-function spreads over the whole space,
any finite value of $Z$ leads to a bound state.

\subsection{Asymptotic solution of the resonant problem}

We now obtain, through the asymptotic method outlined in the paper of
Fernandez and Castro  \cite{Fernandez} an expression for both the ionization rate and
the Stark shift of our model atom. Applying an external and static
electric field to the system, the Schrödinger equation with Hamiltonian
given in the dipole approximation, becomes

\textbf{
\begin{equation}
\left(-\frac{1}{2}\frac{d^{2}}{dx^{2}}-Z\delta(x)-\varepsilon-Fx\right)\psi_{\varepsilon}(x)=0.\label{eq:SchroF}
\end{equation}
}

For $F>0$, the strength of the electric field, and $x\neq0$ this
equation can be written in a more convenient form under the change
of variables $y=\left(2/F^{2}\right)^{1/3}(Fx+\varepsilon)$. It then
becomes the Airy equation with solutions given by the Airy functions
$Ai(-y)$ and $Bi(-y)$. These functions oscillate for $y>0$ and
are exponentially decaying or growing functions for $y<0$, in the
case of $Ai(-y)$ and \textbf{$Bi(-y)$,} respectively. This corresponds,
as expected, to a quasi-bound state brought on by the deformation
of the potential caused by the application of the electric field.
In other words, the application of a static field results in the previously
bound electron having a probability of undergoing tunneling ionization.
Fortunately, a general solution to this Airy equation exists and is
given by

\textbf{
\begin{equation}
\begin{cases}
aAi(-y[x])+bBi(-y[x]) -a<x<0,\\
cAi(-y[x])+dBi(-y[x]) x>0,
\end{cases}
\end{equation}
}where $a$, $b$, $c$, and $d$ are constant coefficients needed
to be dermined.

The wave function that solves equation (\ref{eq:SchroF}) will be
a sum of both Airy functions for $-a<x<0$ in such a way that the
boundary condition for the impenetrable barrier will still be obeyed.
If we define $\beta=y(-a)$ we have that $b=-aAi(-\beta)/Bi(-\beta)$.
We can also impose that the wave-function have an asymptotic form
corresponding to a traveling wave propagating outwards from the origin.
To impose such a condition we look at the asymptotic form of the Airy
functions as $y\to\infty$:

\begin{eqnarray}
Ai(-y)=\pi^{-1/2}y^{-1/4}\sin\left(\frac{2}{3}(-y)^{3/2}+\frac{\pi}{4}\right) && ,\\
Bi(-y)=\pi^{-1/2}y^{-1/4}\cos\left(\frac{2}{3}(-y)^{3/2}+\frac{\pi}{4}\right) && .
\end{eqnarray}

This leads to the choice of coefficients $c=id$,
made in keeping with the idea that the wave-function should look like
an outgoing wave, or in this case, like  $Ci(-y)=Bi(-y)+iAi(-y)$.
We now can turn our attention to the boundary conditions at $y=0$.
These are given by a system of equations consisting of the continuity
of the wave function and discontinuity of its derivative. Defining
$\alpha=y(0)$, $\gamma=Bi(-\beta)/Ai(-\beta)$ we can write boundary
conditions as

\begin{equation}
a(\gamma Ai(-\alpha)-Bi(-\alpha))+cCi(-\alpha)=0
\end{equation}

\begin{eqnarray}
a(2F)^{1/3}\left(Bi'(-\alpha)-\gamma Ai'(-\alpha)\right)\nonumber \\
+2ZcCi(-\alpha)-(2F)^{1/3}Ci'(-\alpha)  =0.
\end{eqnarray}

We can view the previous system as a product of a $2\times2$ matrix
with a vector of coefficients $a$ and $c$ being equal to the zero
vector. For such a system to have non-trivial coefficients $a$ and
$c$ and, thus, a non-trivial wave-function we must have that the
determinant of the matrix be zero. This condition can be written explicitly
as

\begin{eqnarray}
(2Z\pi)Ci(-\alpha)(Ai(-\alpha)Bi(-\beta)\nonumber \\
-Bi(-\alpha)Ai(-\beta))-(2F)^{1/3}Ci(-\beta) =0,\label{eq:DetZero}
\end{eqnarray}
where we have use the Wronskian of the Airy functions, which reads
$Ai'(-\alpha)Bi(-\alpha)-Bi'(-\alpha)Ai(-\alpha)=1/\pi$. We will
now use this result to obtain an asymptotic result for both the Stark
shift and the ionization rate. We start by noting that both $\alpha$
and $\beta$ go to negative infinity in the weak field limit $F\ll1$.
As such, we may replace in equation (\ref{eq:DetZero}) the asymptotic
expressions for the Airy functions when their arguments go to positive
infinity (see Ref.  \cite{Abramowitz}).

\begin{eqnarray}
Ai(y) & =&\frac{1}{2}\pi^{-1/2}y^{-1/4}e^{-\zeta}\sum_{i=0}^{\infty}(-1)^{k}c_{k}\zeta^{-k},\\
Bi(y) & =&\pi^{-1/2}y^{-1/4}e^{\zeta}\sum_{i=0}^{\infty}c_{k}\zeta^{-k},
\end{eqnarray}
where $\zeta=2y^{3/2}/3$, $c_{0}=1$ and $c_{k}=\frac{\Gamma(3k+1/2)}{(54^{k}k!\Gamma(k+1/2)}$.

Setting both real and imaginary parts equal to zero we may obtain
expressions for the real and imaginary parts of the wave vector $k$.
We may notice then, that the imaginary part is small when compared
to the real part, as it can be shown to depend on $e^{-2\zeta}$.
For all intents and purposes, this allow us to write \textbf{$k\approx Re(k)$}.
We may calculate the real part of $k$, for which only the lowest
order terms in $F$ in both series of the Airy functions are considered.
Expanding this in powers of $a$ and equating the result with $k$
itself, and yet again retaining only the lowest order terms, we may
obtain a polynomial equation for $k$ which depends on $F$. Solving
this polynomial equation allows us to write an asymptotic expression
for $Re(k)$ in the limit $F\ll1$, $a\ll1$ as a power series of
$F$. If we write the power series of the energy using $\varepsilon=-k^{2}/2$,
the zeroth order term in $F$ is simply given by the bound state energy
$\varepsilon_{b}=-k_{b}^{2}/2$ and the second order term of this
series corresponds to the Stark shift in the weak field and small
distance limit. Thus we can write the Stark shift $\Delta\varepsilon^{(2)}$
as:

\begin{equation}
\Delta\varepsilon^{(2)}=-\frac{5F^{2}}{8k_{b}^{4}}\label{eq:StarkAssimptotic}
\end{equation}

We can then substitute the bound state energy considering only a zeroth
order approximation for $k$ in the equation for $Im(k)$ and obtain
the imaginary part of the energy. In atomic units the ionization rate
is given by $\Gamma=-2Im(\varepsilon)$ and can, as a result of our
calculations, be written as

\begin{equation}
\Gamma=\frac{k_{b}^{3}}{Z}e^{-\frac{2k_{b}^{3}}{3F}}.\label{eq:IonAssimptotic}
\end{equation}

The presence of $F$ in the denominator of the exponential indicates
that this result is non-perturbative but we note again that the results
of equations (\ref{eq:StarkAssimptotic}) and (\ref{eq:IonAssimptotic})
are only strictly true in the weak field and small distance limit,
and as such, in the following sections we present an alternative exact
method for calculating the Stark shift valid for all values of $a$
and therefore characterize the response of the model atom to an applied
electric fields through an exact calculation of the static polarizability.

\section{Exact calculation of the static and dynamic polarizability\label{sec:Exact-calculation-of}}

Usually, the direct approach to the calculation of a response function
faces enormous difficulties due to the necessity of summing over all
eigenstates of the system. Dalgerno and Lewis \cite{Dalgarno} circumvented this difficulty
reducing the need of summing over all states to the solution of an
inhomogeneous differential equation. In this section we obtain the
exact expression to the static polarizability of our model-atom using
the Dalgarno-Lewis approach. This same approach is possible in the
case of a frequency dependent electric field using a method due to Fowler
\cite{Fowler}. The Dalgarno-Lewis method (and its generalization
to time-dependent problems performed in \cite{Karplus1962,karplus_1963})
has been applied to many different problems, relevant examples being
the effect of phonons on the ground state of a Wannier exciton discussed
in \cite{Vooght_1973}; the effect of confinement on the Stark effect
analyzed by Pedersen \cite{Pedersen_2017}; the study of nonlinear optics
considered in \cite{Thayyullathil_2003,Radhakrishnan_2004}; and
the list goes on, including particle physics and astrophysics.  We
believe that this method may also be useful for students first learning
about perturbation theory methods, as its discussion not only shines
a spotlight on the difficulty of evaluating infinite sums over states,
but also, provides an easily graspable and elegant workaround to this
problem. Here we use the method to compute both the static and the
dynamic polarizability of our semi-confined system as an illustration
of its usefulness as well as a means of capturing the physics of the
simple model we proposed.

\subsection{The Dalgarno-Lewis method revisited}

The Dalgarno-Lewis perturbation theory is seldom
discussed in text books on quantum mechanics. This is very unfortunate
since the method is very powerful for summing an infinite number of
states (a task that cannot, in general, be done by a direct brute-force
approach). Some notable exceptions which do cover Dalgarno-Lewis theory
are the books of Desai \cite{desai_2009} which uses the method to calculate
the polarizability of an Hydrogen atom, as well as  Refs.  \cite{konishi2009quantum}
and  \cite{Schiff} which provide a pedagogical account of the theory.
Therefore, we give here a brief, but enough detailed, derivation of
Dalgarno-Lewis perturbation theory, using Dirac notation for generality
and following the approach of Schiff \cite{Schiff} .    Some knowledge of Rayleigh-Schrodinger
(RS) perturbation theory is assumed for this section, such as the
form of the second order correction to the energy, or the first order
correction to the wave-function, but accounts of these topics can
be found in most classic textbooks on quantum mechanics. However,
we note that there exist approaches to the derivation of Dalgarno-Lewis
theory that do not rely so much on usual RS perturbation theory, such
as that developed in \cite{balantekin2010dalgarno} for algebraic
Hamiltonians.

It is a well known result from RS perturbation theory that if a Hamiltonian
can be written in the form $H=H_{0}+H'$, where $H'$ is a perturbation
that is small compared to $H_{0}$, then the second order correction
to the energy for a non-degenerate state (in particular the ground-state
of a one dimensional system, which is always non-degenerate) is given
by

\begin{equation}
\Delta\varepsilon^{(2)}=\sum_{n\neq m}\frac{\left\langle m\left|H'\right|n\right\rangle \left\langle n\left|H'\right|m\right\rangle }{E_{m}-E_{n}}\label{eq:Sum}
\end{equation}
where the sum runs over both the remaining bound states and the whole
continuum of scattering states. In the case that the states form a
continuum, this is to be replaced by an integral. The Dalgarno-Lewis
method starts by assuming that there exists an operator $O$ such
that

\begin{equation}
\frac{\left\langle n\left|H'\right|0\right\rangle }{E_{0}-E_{n}}=\left\langle n\left|O\right|0\right\rangle .\label{eq:defO}
\end{equation}
This might seem, at first, like a large leap or arbitrary choice,
but we shall see that the operator $O$ has a precise physical meaning.
Writing the second order correction to the energy using the operator
$O$ we have

\begin{eqnarray}
\Delta\varepsilon^{(2)} & =&\sum_{n\neq0}\left\langle 0\left|H'\right|n\right\rangle \left\langle n\left|O\right|0\right\rangle \nonumber \\
 & =&\left\langle 0\right|H'\sum_{n\neq0}\left|n\right\rangle \left\langle n\right|O\left|0\right\rangle \label{eq:shift}
\end{eqnarray}
The sum of $\left|n\right\rangle \left\langle n\right|$ over the
whole Hilbert space is a partition of the identity and, since in equation
(\ref{eq:shift}) the sum runs over all but the ground state, we may
write

\begin{equation}
\Delta\varepsilon^{(2)}=\left\langle 0\left|H'O\right|0\right\rangle -\left\langle 0\left|H'\right|0\right\rangle \left\langle 0\left|O\right|0\right\rangle .
\end{equation}

We have thus avoided the integral over states. We have only to rewrite
this equation in such a way as to bring out the dependence in the
first order correction to the wave-function. To do this we make use
of the definition of $O$ and are able to write $\left\langle n\left|H'\right|0\right\rangle $
as $\left\langle n\left|[O,H_{0}]\right|0\right\rangle $ where $[,]$
is the commutator. This can be easily checked by simply letting $O$
act on $\left|0\right\rangle $ and then rewriting the energy eigenvalues
as the operator $H_{0}$ acting on the respective states:

\begin{eqnarray}
\left\langle n\left|H'\right|0\right\rangle  & =&\left\langle n\left|O\left(E_{0}-E_{n}\right)\right|0\right\rangle \nonumber \\
 & =&\left\langle n\right|-H_{0}O+OH_{0}\left|0\right\rangle \nonumber \\
 & =&\left\langle n\right|[O,H_{0}]\left|0\right\rangle 
\end{eqnarray}

This equality however can be written as an equality between operators
up to a constant $[O,H_{0}]=H'+C$.  The equality
holds for an arbitrary choice of $C$, since, by non-degenerate perturbation
theory, the sum occurs over $n\neq0$ and the states are orthogonal,
from which it is clear that $\left\langle n\right|C\left|0\right\rangle =C\left\langle n\right.\left|0\right\rangle =0$.
In particular, taking $n=0$, we can now say that

\begin{equation}
\left\langle 0\left|[O,H_{0}]\right|0\right\rangle =\left\langle 0\left|H'\right|0\right\rangle +C.
\end{equation}
We now take advantage of the freedom in the choice of $C$. Since
it must be true tha t $\left\langle 0\left|[O,H_{0}]\right|0\right\rangle =0$
we then have that $C=-\left\langle 0\left|H'\right|0\right\rangle $.
If we now define $\left|\psi_{1}\right\rangle =O\left|0\right\rangle $
we may write

\begin{equation}
(E_{0}-H_{0})\left|\psi_{1}\right\rangle =H'\left|0\right\rangle -\left\langle 0\left|H'\right|0\right\rangle \left|0\right\rangle .\label{eq:DNequation}
\end{equation}
From Rayleigh-Schrödinger Perturbation theory it is easy to show that
$\left|\psi_{1}\right\rangle $ corresponds, by definition of the
operator $O$, to the first order correction to the wave function.
We have only to multiply on the left and right hand sides of equation
\ref{eq:defO}, by $\left|n\right\rangle $ and then sum over $n\neq0$.
Using the partition of the identity, we are left with the usual expression
for the first-order correction:

\begin{eqnarray}
\sum_{n\neq0}\frac{\left\langle n\left|H'\right|0\right\rangle }{E_{0}-E_{n}}\left|n\right\rangle  & =&\sum_{n\neq0}\left(\left|n\right\rangle \left\langle n\right|\right)\left|\psi_{1}\right\rangle \nonumber \\
&\Leftrightarrow & \left|\psi_{1}\right\rangle =  \sum_{n\neq0}\frac{\left\langle n\left|H'\right|0\right\rangle }{E_{0}-E_{n}}\left|n\right\rangle .\label{eq:sumwave}
\end{eqnarray}

We now see that the operator $O$ gives the first
order correction to the wave-function when acting on the ground state,
and can be regarded as the first term $O_{0,1}$ in a series $S_{n}=\sum_{i}O_{n,i}$
that gives the eigenstates of the full Hamiltonian $H$ from the eigenstates
$\left|n\right\rangle $ of the unperturbed Hamiltonian $H_{0}$ by
generating the successive corrections (see Ref.  \cite{balantekin2010dalgarno}).

From these considerations, we can see that the Dalgarno-Lewis procedure
amounts to substituting the sum over states of equation (\ref{eq:Sum}
or \ref{eq:sumwave}) with the solution of the inhomogeneous differential
equation (\ref{eq:DNequation}), referred to as the Dalgarno-Lewis
equation. This solution gives us the first order correction to the
wave function, while from usual perturbation theory, the correction
to the energy is evaluated as

\begin{equation}
\Delta\varepsilon^{(2)}=\left\langle 0\left|H'\right|\psi_{1}\right\rangle .\label{eq:final}
\end{equation}
In the rest of the paper, this method will be applied in the next
section to the calculation of the Stark shift and static polarizability
of the model-atom.

\subsection{Application to the calculation of the static electric polarizability}

The static polarizability is a response function of a system to a
static electric field. For non-degenerate ground states the linear
order polarizability is zero. The second order polarizability, however,
is finite and it will concern us in this section. Let us apply the
formalism introduced in the previous subsection, now adapted to a
problem written in real space. Our bound state $\left|0\right\rangle $
corresponds to the bound state wave function given in equation (\ref{eq:gs}),
the zeroth order Hamiltonian $H_{0}$ will have the potential included
in equation (\ref{eq:schro}) and the perturbation Hamiltonian $H'$
is given by the term of equation (\ref{eq:SchroF}) that depends on
the electric field intensity $-Fx$. The Dalgarno-Lewis equation reads,
in real space

\begin{eqnarray}
\left(-\frac{k_{b}^{2}}{2}+\frac{1}{2}\frac{d^{2}}{dx^{2}}+Z\delta(x)\right)\psi_{1}(x)\nonumber \\
=\left(-Fx-\int_{-a}^{\infty}dx\ (-Fxu_{b}^{2}(x))\right)u_{b}(x).\label{eq:DNeq2}
\end{eqnarray}

Both the integral and the resulting differential equation can be solved
as shown in the Appendix. The correction to the wave function is obtained
as explained in the Appendix, and the integral entering in equation
(\ref{eq:DNeq2}) can be readily computed yielding

\begin{equation}
\Delta\varepsilon^{(2)}=\frac{(-2a(2a^{2}k_{b}^{2}-15)(Z-k_{b})-15)}{24k_{b}^{4}(2a(k_{b}-Z)+1)}F^{2}.\label{eq:StarkShift}
\end{equation}

This result is, in contrast to those presented in Eq. (\ref{eq:StarkAssimptotic}),
valid for all values of strength of the atom potential $Z$ as well
as the distance $a$. In particular, when $a\ll1$ if we retain only
powers of zero order of $a$ we recover the asymptotic result previously
derived for the Stark shift.

\begin{figure}[h]
\begin{centering}
\includegraphics[scale=0.4]{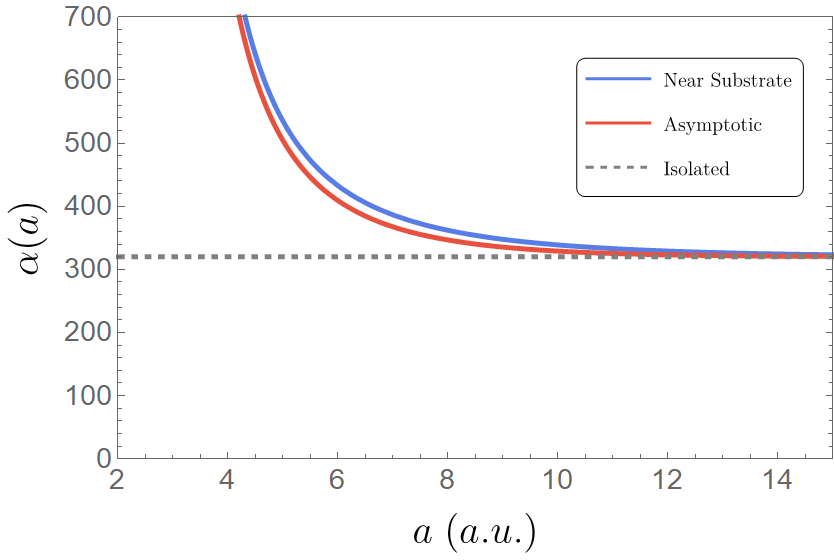}
\par\end{centering}
\caption{Plot of the static polarizatbility of the isolated model-atom (dashed
and gray) and model-atom in the vicinity of a substrate (blue). We
compare the results with the asymptotic calculation given in equation
(\ref{eq:StarkAssimptotic}) (red). Here we have chosen the effective
nuclear charge $Z=0.25$ and plot the polarizability as a function
of distance from the substrate, $a$, in atomic units (a.u.). Note
that in our atomic system modeled by a $\delta$-function potential,
there is no\textit{ a priori} restriction that $Z$ be whole and thus
can take values smaller than 1. As can be seen, the presence of the
substrate causes an extremely sharp increase in polarizability for
small distances, which are close to the limiting condition $a=1/(2Z)$
obtained from equation (\ref{eq:ktrans}). As is expected, for larger
distances, the effects of the presence of the substrate are much smaller
and both polarizabilities coincide for large $a$. The value $Z=0.25$
is chosen so that there occurs significant deviation from the polarizability
of the isolated atom for $a$ at the scales below 10 a.u. or, in S.I.
units, at 0.5 nm.\label{fig:Polarizability-as-a}}
\end{figure}

We now turn our attention to an interpretation of the Stark shift
as a measure of the second order static polarizability, which corresponds
to an induced dipole moment measured by the static polarizability
$\alpha$. The polarizability is itself defined through

\begin{equation}
\Delta\varepsilon^{(2)}=-\frac{1}{2}\alpha F^{2}.\label{eq:Polariz}
\end{equation}
Thus, from equation (\ref{eq:StarkShift}) we can see that the static
polarizability is given by

\begin{equation}
\alpha=\frac{(2a(2a^{2}k_{b}^{2}-15)(Z-k_{b})+15)}{12k_{b}^{4}(2a(k_{b}-Z)+1)}.
\end{equation}
If we hold $Z$ fixed as a parameter of our atomic model, we may vary
the distance $a$ and see how this affects the static polarizability.
Note that $k_{b}$ will also change as a consequence of the transcendental
equation (\ref{eq:ktrans}). The polarizability will increase dramatically
as a result of the presence of the substrate located at $x=-a$ as
can be seen in Fig. \ref{fig:Polarizability-as-a} in blue, and will
otherwise tend to a finite value corresponding to the static polarizability
of the isolated atomic system as the distance from substrate becomes
large. From Fig. \ref{fig:Polarizability-as-a} we notice a strong
deviation from the isolated atom polarizability for values of $a$
smaller than 10 a.u.$=$0.5 nm. Considering that the size of a carbon
atom is about 0.22 nm, the change in the static polarizability will
be sensitive to differences in one layer to two layers of atoms separating
the atomic system from the substrate, as depicted in Fig. \ref{fig:Schematic-illustration-of}.

\subsection{Calculation of the dynamic polarizability}

In this section we concern ourselves with the application of an external
time and frequency dependent electric field and therefore study the
optical response of the system,   thus showing that
the Dalgarno-Lewis method can also be applied to time-dependent problems.
This amounts to considering an applied electric field of the form
$F(t)=-F(e^{i\omega t}+e^{-i\omega t})$ and solving the time dependent
Schrodinger equation:

\begin{equation}
i\frac{\partial}{\partial t}\psi_{\varepsilon}(x,t)=\left[H_{0}+F(t)x\right]\psi_{\varepsilon}(x,t),\label{eq:TimeDep}
\end{equation}
where $H_{0}$ is the Hamiltonian of the atomic system in proximity
to the substrate. Following the approach of Fowler we consider a time
dependent perturbation such that:

\begin{eqnarray}
\psi_{\varepsilon}(x,t) && \approx u_{b}(x)e^{ik_{b}^{2}t/2}+\phi_{1}^{+}(t)e^{i(k_{b}^{2}/2+\omega)}\nonumber \\
 && +\phi_{1}^{-}(t)e^{i(k_{b}^{2}/2-\omega)}.\label{eq:TimeDepPert}
\end{eqnarray}

Substituting equation (\ref{eq:TimeDepPert}) into equation (\ref{eq:TimeDep})
and grouping terms in the same exponentials we further obtain a differential
equation for $\phi_{1}^{\pm}$ of the form:

\begin{equation}
\left[-\frac{1}{2}\frac{\partial^{2}}{\partial x^{2}}-Z\delta(x)+\left(\frac{k_{b}^{2}}{2}\pm\omega\right)\right]\phi_{1}^{\pm}(x)=Fxu_{b}(x).\label{FowlerEq}
\end{equation}

As before, we are able to determine the first order correction to
the wave-function, as well as the Stark-shift through the use of equation
(\ref{eq:final}) and which now takes the form

\begin{equation}
\alpha(\omega)=\frac{2}{F^{2}}\left[\left\langle b\right|Fx\left|1\right\rangle ^{+}+\left\langle b\right|Fx\left|1\right\rangle ^{-}\right].\label{eq:polarizdef}
\end{equation}

\begin{figure}[h]
\begin{centering}
\includegraphics[scale=0.39]{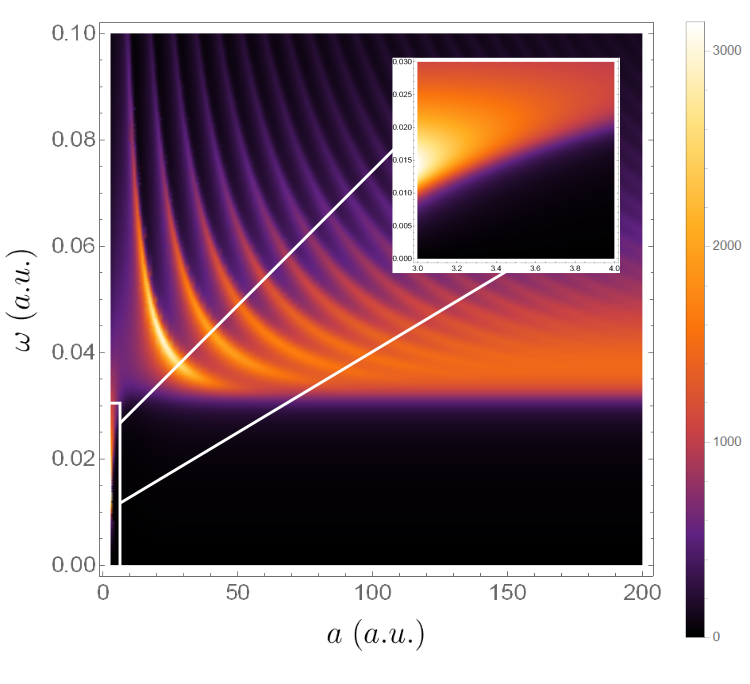}
\par\end{centering}
\caption{Plot of the imaginary part of the dynamic polarizatbility of the atomic
system near a substrate. Here we have chosen the effective nuclear
charge $Z=0.25$ and plot the polarizability as a function of distance
from the substrate and frequency in atomic units. Firstly we note
that there exist peaks of greater polarizablity in the vicinity of
the substrate. We also find that there exists a very sharp increase
of polarizability for small frequency and small distance (see zoomed
in portion) as is necessary for the dynamic polarizability to match
the static case as $\omega\to0$. For large distances the peaks and
troughs of polarizability meld together and converge to the result
of the isolated atom visible on the bottom right as a continuous slope.
Note that a small imaginary part was added to the frequency for the
results to be finite (the value of the imaginary part used is 0.0018
a.u.$=$50 meV). This is necessary as it accounts for a broadening
of the energy levels of the atomic system caused by the existence
of a decay rate associated with ionization.\label{fig:DynamicPolariz}}
\end{figure}

\begin{figure}
\begin{centering}
\includegraphics[scale=0.4]{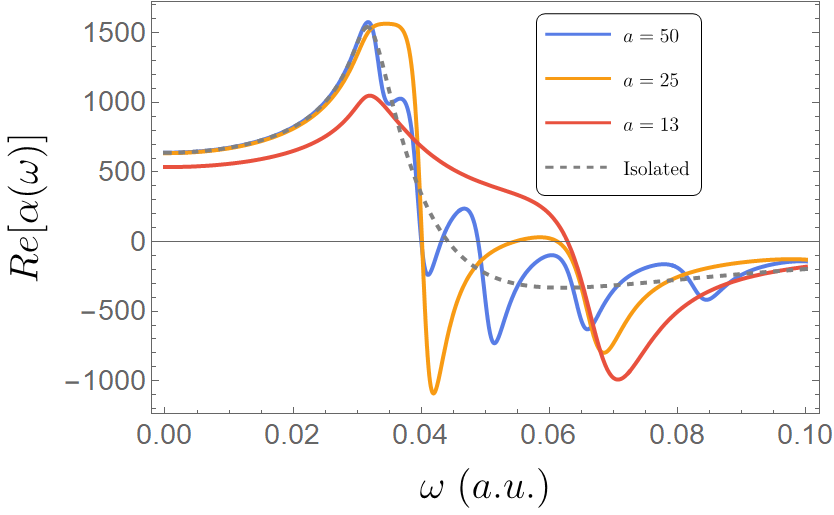}
\par\end{centering}
\begin{centering}
\includegraphics[scale=0.4]{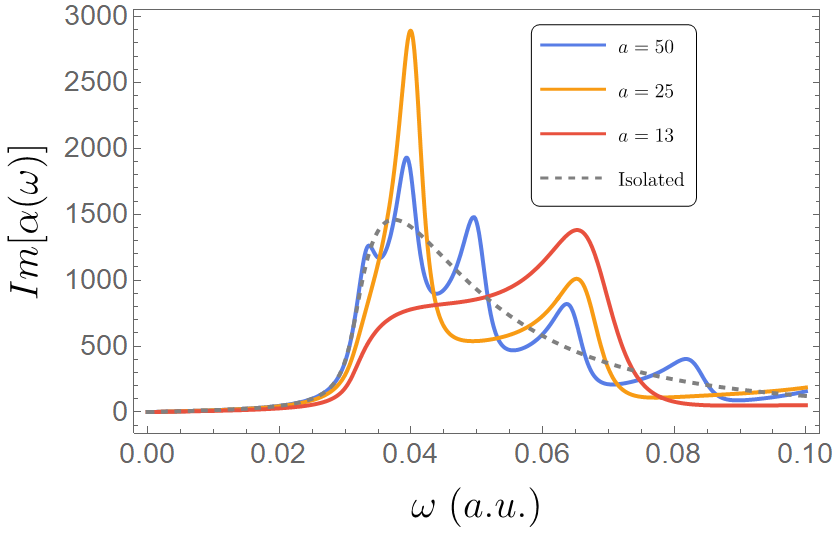}
\par\end{centering}
\caption{Plot of the real and imaginary parts of the dynamic polarizability
of the atomic system with $Z=0.25$ as a function of frequency of
the applied field for several values of distance $a$ (blue, orange
and red). These curves can be understood as vertical cuts of Fig.
\ref{fig:DynamicPolariz}. We also plot the dynamic polarizability
of the isolated atom (gray and dashed) as a reference for the effects
of the presence of the substrate. From the plots we see that decreasing
$a$ results in an increase in the size of the resonances of the polarizability,
but also a decrease in the number of these resonances. Further increasing
the distance will cause the peaks to occur with a much higher frequency
but be smaller in size and the curve of the polarizability will eventually
be indistinguishable from that of the isolated atom. This is also
visible in Fig. \ref{fig:DynamicPolariz}. As in the latter, a small
imaginary part ( 0.0018 a.u.$=$50 meV) was added to the frequency
for the imaginary part of polarizability to be finite. \label{fig:DynamicPolarizA}}
\end{figure}

Where $\left\langle x|1\right\rangle ^{\pm}=\phi_{1}^{\pm}(x)$. This
allows us to obtain an analytic expression for the dynamic polarizability
of the model-atom through the use of equation (\ref{eq:Polariz}),
which we omit due to it being extremely cumbersome. In essence, this
approach is very similar to the Dalgarno-Lewis method in the sense
that an infinite sum over states is also avoided through the solution
of an inhomogeneous differential equation, however a dependence on
the frequency comes into play, which as seen in Fig. \ref{fig:DynamicPolariz}
and \ref{fig:DynamicPolarizA} plays an important role in the physical
description of the system.

In Fig. \ref{fig:DynamicPolariz} we see that despite the fact that
the imaginary part of the dynamic polarizability is nonmonotonic in
$a$, the size of its peaks become larger as $a$ becomes smaller
We therefore are able to conclude that, electromagnetic hot-spots
can occur when the system is subjected to a frequency-dependent electric
field even for large distances to a substrate, but in particular,
these effects are greatly exacerbated for small distances to the substrate.
More importantly, the shape of the curves of the polarizability as
function of frequency are considerably different for different distances
of the atomic system to the substrate. Therefore, scanning the system
in frequency and retrieving its dynamic polarizability gives a measure
of the distance of the atomic system to the substrate or, in the context
of probing a deposit on the substrate, a measure of the number of
molecular layers in between the substrate and atomic system. The discrimination
of the distance can extend up to $a=50$ a.u.. We also note the existence
of strong oscillations in the imaginary part of $\alpha(\omega)$
for large values of $a$. These are due to interference effects brought
by the presence of the impenetrable substrate, originated on the scattering
states.

\section{Conclusion and outlook\label{sec:Conclusion-and-outlook}}

In this paper we have studied several aspects of the electrical and
optical response of a one dimensional atomic system consisting of
a $\delta$-function when close to a substrate and subjected to an
external and static as well as a frequency-dependent electric field.
This was done in the context of providing a simple and physically
relevant example of the application of the Dalgarno-Lewis method in
perturbation theory. We started by giving a limiting condition for
the existence of a bound state and obtaining a transcendental equation
that allows for the calculation of the bound state energy as a function
of distance and of the effective nuclear charge of our toy-model.
We then provided asymptotic results for the ionization rate and Stark
shift using asymptotic expansions of the Airy functions and the boundary
conditions of our problem in order to write a condition for a non-vanishing
wave function. The limit for small distances of such an expression
allows for the calculation of a complex energy of which the real part
corresponds to the Stark shift and imaginary part to the ionization
rate. We have also provided a general derivation of the Dalgarno-Lewis
method for perturbation theory, which we believe could be a welcome
addition to many advanced undergraduates and early graduate students'
tool-belts. The method allowed for a calculation of an exact second
order correction to the energy for any distance to the substrate,
which would otherwise be difficult for such a geometry. This result
was seen to match the asymptotic one in the appropriate limit and
allowed for a characterization of the static polarizability of our
model-atom as a function of distance to the substrate, and unraveled
some rich behavior within our simple model: The polarizability was
seen to match the isolated atom for large distances, as expected,
but we showed that it increases quite dramatically when the atomic
system is close to the substrate. Therefore, we argue that the measurement
of the static polarizability of atomic systems can act as a ``molecular
ruler'', useful, for instance, in the context of the measurement
of atomic distances to a substrate. An example can be the measurement
of the thickness of atomic layers, deposited on a substrate. Through
the calculation of the dynamic polarizability we have showed the existence
of large resonances in the dynamic polarizability for small distances,
which in the context of a substrate bound to DNA or other organic
molecules, may result in an increased Raman-scattering signal from
these molecules to which the atomic system is bound. This may therefore
allow for an increased sensitivity in the detection of the presence
of organic molecules bound to surfaces and covered by polarizable
particles. The shape of the curve of the dynamic polarizability as
a function of the applied electric field frequency was shown to be
very sensitive to the distance to the substrate. As such we have argued
that the change of the polarizability with the distance, which can
be rather strong, can function as a nanometric ruler both in the static
and dynamic cases and for small enough distances. For the parameters
shown in the figures the strong enhancement occurs in the scale below
1 nm for the static polarizability but we may also consider the case
of an atomic system with a smaller effective charge $Z$ where the
enhancement occurs at larger scales. In the case presented in the
text, the system would indeed function as an atomic ruler, allowing
to probe very thin layers of deposited molecules on a substrate. In
particular, we can envision a strong interaction between the atomic
system and the panoply of recently obtained two-dimensional (2D) materials.
These systems, being essentially a surface are the perfect test bed
of our ideas, since we would have both a very small distance between
the 2D-materials and high sensitivity to small changes in the distance
between the 2D-material and the atomic system.

\ack

The authors acknowledge João Carlos Henriques and Thomas G. Pedersen
for many discussions and exchange of ideas on this topic. N. M. R.
P acknowledges support from the European Commission through the project
``Graphene-Driven Revolutions in ICT and Beyond'' (Ref. No. 785219),
and the Portuguese Foundation for Science and Technology (FCT) in
the framework of the Strategic Financing UID/FIS/04650/2019. In addition,
he acknowledges COMPETE2020, PORTUGAL2020, FEDER and the Portuguese
Foundation for Science and Technology (FCT) through projects POCI-01-0145-FEDER-028114,
POCI-01-0145-FEDER-029265, PTDC/NAN-OPT/29265/2017, and POCI-01-0145-FEDER-02888.
T. V. C. A. acknowledges FCT for a grant given in the context of the
summer training program 2020: ``Quantum Matter | Materials \& Concepts
Summer Training Program 2020''.

\appendix

\section{Solving the Dalgarno-Lewis Equation}

We show, in this appendix, a few steps of the solution of the Dalgarno-Lewis
equation (\ref{eq:DNeq2}). We start by solving the Integral

\begin{eqnarray}
\int_{-a}^{\infty}dx\ Fx\ u_{b}(x) && =2Ae^{-k_{b}a}\int_{-a}^{0}dx\ \sinh(k_{b}(x+a))\nonumber \\
 & &+2Ae^{-k_{b}a}\sinh(k_{b}a)\int_{0}^{\infty}dx\ e^{-k_{b}x}.
\end{eqnarray}

By direct computation of both integrals we obtain a simple coefficient

\begin{equation}
\int_{-a}^{\infty}dx\ Fx\ u_{b}(x)=\frac{aF(Z-k_{b})}{2a(Z-k_{b})-1}.
\end{equation}

Where we have also used the transcendental equation for $k_{b}$.
This coefficient is merely a constant with respect to $x$ which we
will now call $D$. The Dalgarno-Lewis equation therefore becomes,
by virtue of the piece-wise definition of the bound state wave-function,
a piece-wise, second order inhomogeneous differential equation defined
as

\begin{eqnarray}
\psi_{1}''(x)-k_{b}^{2}\psi_{1}(x)  =2(Fx-D)2Ae^{-k_{b}a}\times\nonumber\\
\sinh(k_{b}(x+a))\,,  \hspace{0.5cm}-a<x<0,\\
\sinh(k_{b}a)e^{-k_{b}x}\,,   \hspace{0.5cm} x>0.
\end{eqnarray}

Both parts of the equation consist of inhomogeneous differential equations
where the term independent of $\psi_{1}(x)$ is either an exponential,
a polynomial or at most a product of the two. There are many well
known methods for solving such equations, and applying any of them
yields a solution for $\psi_{1}(x)$. We omit here this function as
its analytic representation is cumbersome and brings no practical
insight despite consisting only of products of exponential functions
and polynomials. The energy correction corresponding to the Stark
shift requires the calculation of yet another integral corresponding
to plugging in $\psi_{1}(x)$ as well as $-Fx$ into equation (\ref{eq:final})

\begin{equation}
\Delta\varepsilon^{(2)}=-F\int_{-\infty}^{\infty}dx\ u_{b}(x)x\psi_{1}(x).
\end{equation}

This is yet again an integral of products of exponentials and polynomials
and, as such, a computation of relative ease yields equation (\ref{eq:StarkShift}).

\section*{References}

\bibliographystyle{iopart-num}
%

\end{document}